# Integrating Large Language Models with Graphical Session-Based Recommendation

NAICHENG, GUO, MYbank, Ant Group, China
HONGWEI, CHENG, MYbank, Ant Group, China
QIANQIAO, LIANG, MYbank, Ant Group, China
LINXUN, CHEN, MYbank, Ant Group, China
BING, HAN, MYbank, Ant Group, China

With the rapid development of Large Language Models (LLMs), various explorations have arisen to utilize LLMs capability of context understanding on recommender systems. While pioneering strategies have primarily transformed traditional recommendation tasks into challenges of natural language generation, there has been a relative scarcity of exploration in the domain of session-based recommendation (SBR) due to its specificity. SBR has been primarily dominated by Graph Neural Networks, which have achieved many successful outcomes due to their ability to capture both the implicit and explicit relationships between adjacent behaviors. The structural nature of graphs contrasts with the essence of natural language, posing a significant adaptation gap for LLMs. In this paper, we introduce large language models with graphical Session-Based recommendation, named LLMGR, an effective framework that bridges the aforementioned gap by harmoniously integrating LLMs with Graph Neural Networks (GNNs) for SBR tasks. This integration seeks to leverage the complementary strengths of LLMs in natural language understanding and GNNs in relational data processing, leading to a more powerful session-based recommender system that can understand and recommend items within a session. Moreover, to endow the LLM with the capability to empower SBR tasks, we design a series of prompts for both auxiliary and major instruction tuning tasks. These prompts are crafted to assist the LLM in understanding graph-structured data and align textual information with nodes, effectively translating nuanced user interactions into a format that can be understood and utilized by LLM architectures. Extensive experiments on three real-world datasets demonstrate that LLMGR outperforms several competitive baselines, indicating its effectiveness in enhancing SBR tasks and its potential as a research direction for future exploration.

## 1 INTRODUCTION

Recommender systems have become an integral part of various platforms, recommending relevant resources to users based on their preferences. However, there are realistic scenarios where only limited interactions are

Authors' addresses: NAICHENG, GUO, guonaicheng.gnc@mybank.cn, MYbank, Ant Group, Beijing, Beijing, China; HONGWEI, CHENG, MYbank, Ant Group, Shanghai, China; QIANQIAO, LIANG, MYbank, Ant Group, Shanghai, China; LINXUN, CHEN, MYbank, Ant Group, Beijing, China; BING, HAN, MYbank, Ant Group, Shanghai, China.

available to detect user intent, making traditional recommenders inadequate. To address this, researchers proposed session-based recommendation (SBR), which relies on user behavior sequences for recommendations. Traditional SBR methods use Markov Chains [10, 28], Convolution Neural Networks (CNNs) [30], Recurrent Neural Networks (RNNs) [11, 19], and attention mechanism [25], to capture sequential signals and dynamic user interests. Newly, Graph Neural Networks (GNNs) have been incorporated into SBR, leveraging their ability to model structural information and complex transitions between items in a session. GNNs have rapidly emerged as a state-of-the-art (SOTA) approach [6, 36, 39] in SBR tasks, surpassing traditional methods with their impressive ability to capture complex item transitions and relationships within user sessions. The GNN-based methods excel at discerning the intricate patterns of user behavior, thanks to their sophisticated representation of items as nodes within a graph structure, allowing for the accurate modeling of interactions and affinities between items. By capitalizing on this structural information, GNN-based methods have demonstrated superior performance in predicting user preferences in SBR tasks with limited interaction data. Despite significant progress, existing SBR algorithms primarily rely on user interaction data, neglecting valuable textual information associated with users and items. This restricts the algorithms' capability to capture the nuances and context of interactions.

Recently, the emergence of large language models (LLMs), which excel in assimilating real-world knowledge from the Web and achieving proficient natural language generation, has triggered a significant revolution in the research community [43]. Specifically, there are also several attempts [5, 12, 22] that adapt LLMs zero/few-shot ability for recommender systems, which struggle to provide accurate recommendations often due to a lack of specific recommendation task training. Moreover, increasing efforts have predominantly converted the recommendation task into a natural language generation task to further fine-tune LLMs using relevant recommendation data [2, 3, 16, 40] to address the above limits. By integrating LLMs, recommender systems can benefit from the rich textual information associated with users and items [38, 41], tapping into the descriptive content to better infer user preferences and improve the accuracy of the recommendations provided.

Inspired by the success of LLMs in traditional recommendation systems, a natural inquiry arises: Can existing LLMs be applied to the graph-based SBR task? However, SBR approaches heavily rely on graph structures, and can not be directly processed by LLMs, missing out on the benefits of existing LLM-based recommendation methods. To bridge this gap and leverage the capabilities of LLMs, a fundamental problem needs to be addressed for successful item recommendations.

**Challenge: How to express graph-based SBR tasks in natural language?** Graph-based SBR tasks are inherently multi-faceted, encompassing both the representation of complex item interactions and the dynamic nature of user interests. In contrast to other recommendation scenarios where tasks can be readily described as language generation tasks, as shown in Fig. 1, SBR methods use inherently structured graph data, which is not naturally aligned with the sequential and interpretive processing of LLMs. The structural nature of graphs contrasts with the essence of natural language, posing a significant adaptation challenge for LLMs.

To overcome these challenges above, we integrate **L**arge **L**anguage **M**odels with **G**raph-Based Session-Based **R**ecommendation, named LLMGR, which enhances session-based recommenders by flexibly adapting LLM. LLMGR is designed to enhance SBR by exploiting the vast knowledge and semantic understanding capabilities of LLMs, addressing the challenges of capturing user preferences from limited interaction data. Firstly, we engineer a series of prompts that guide the model through behavior pattern modeling and graph comprehension tasks. These prompts combine textual templates with placeholders, facilitating the hybrid encoding of both linguistic and graph-structural information into a format amenable to LLM processing. Secondly, we design a two-stage tuning process: an auxiliary stage focuses on establishing text-node associations, as well as a main stage captures behavioral patterns within session graphs. Finally, to evaluate our approach, we conduct extensive experiments on three real-world datasets, and our method achieves the best performance compared to several competitive baselines. Our goal is not only to enhance the current understanding of session-based recommendations but also to provide pioneer attempts for future research at the intersection of complex graph modeling and advanced language models.

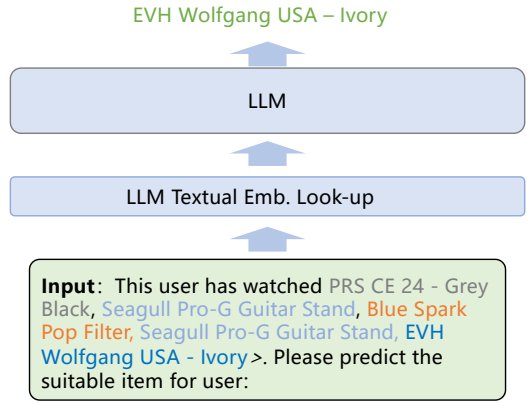

Fig. 1. Illustration of LLM-based method for conventional recommendation scenarios.

Overall, our major contributions can be summarized as follows:

- We design an LLM-based framework that aims at addressing the shortcomings of SBR in effectively utilizing textual information. To our knowledge, this is the first framework to tackle this issue using LLM.
- We tailor an effective two-stage instruction tuning strategy to boost LLM capability to decipher the intricate relationships within session graphs, understand graph structure, and capture the behavior pattern simultaneously.
- Extensive experimental results indicate that our proposed LLMGR not only significantly outperforms the SOTA comparison method but also has portability to benefit the conventional SBR methods.

## 2 PRELIMINARIES

In this section, we introduce the notation and formalize the problem we aim to address. Additionally, we provide a concise overview of SBR methods to set the stage for our subsequent discussions.

### 2.1 Problem Formulation

A session-based recommendation task is constructed on historical user behavior sessions and makes predictions based on current user sessions. In this task, there is an item set $\mathcal{V}$, where $m = |\mathcal{V}|$ is the number of items and all items are unique. Each session $S = [v_1, v_2, \ldots, v_n]$ is composed of a series of user's interactions, where $v_i$ represents an item clicked at time $i$ in $S$ and $n$ represents session's length. SBR task is to predict the item that the user is most likely to click on next time in a given session $S$. For each given session $S$ in the training process, there is a corresponding label $y$ as the target. In the training process, for each item $v_i \in \mathcal{V}$ in a given session, our model learns the corresponding embedding vector $v \in \mathbb{R}_1^d$, where $d_1$ is the dimension of node $v$. Our framework outputs a probability distribution $\hat{y}$ over the given session $S$.

### 2.2 Graph-based SBR Methods

Most recent SOTA SBR methods have utilized GNNs to model the coherence of items within a session graph due to GNNs possess powerful capabilities for representing structured data and capturing implicit dependencies. Fig.

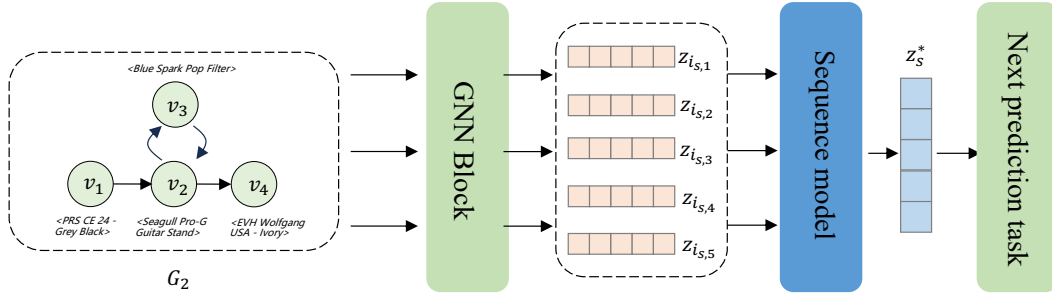

Fig. 2. The overall paradigm of GNN-based Session-based recommendation.

2 illustrates the representative overall paradigm of GNN-based SBR. Graph-based SBR methods can be roughly divided into the following parts: session graph construction, information propagation, and session output.

*2.2.1 Session Graph Construction.* By capturing sequential patterns among adjacent items in a user's behavior sequence, we transform each behavior sequence $S = [v_1, v_2, ..., v_i]$ into a corresponding graph $G = (v, e)$, where $v \in V$ represents the set of vertices and $e$ represents the set of directed edges. An illustration of a session graph built from the sequence $[v_1, v_2, v_3, v_2, v_4]$ is provided in Fig. 2, showcasing edges that connect each pair of consecutive items.

*2.2.2 Information Propagation.* GNNs can effectively capture and model complex relationships and dependencies present in graphs. Existing works aim to learn a classifier and the graph-level representation to predict the label of the graph. Given a collection of graphs $(G_1, G_2, \ldots, G_n) \in G$ and the corresponding labels $(v_{L_1}, v_{L_2}, \ldots, v_{L_n}) \in V_L$. GNNs use the structure of the graph and the original feature of each node to learn its corresponding representation. The learning process is to take a node as the center, and iteratively aggregate the neighborhood information along edges. The information aggregation and update process can be formulated as follows:

$$\mathbf{t}_v^{(l+1)} = f_{\text{aggregator}}(\mathbf{x}_u^{(l)}, u \in N(v)), \tag{1}$$

$$\mathbf{x}_v^{(l+1)} = f_{\text{updater}}(\mathbf{x}_v^{(l)}, \mathbf{t}_v^{(l+1)}), \tag{2}$$

where $\mathbf{x}_v^l$ represents the embedding of node $v$ after $l$-th layer aggregator and $N(v)$ is neighborhood of node $v$. The information aggregation function $f_{\text{aggregator}}$ aggregates the information from the neighborhood information and passes it to the target $v$. The update function $f_{\text{updater}}$ calculates the new node statues from the source embedding $\mathbf{x}_v^l$ and the aggregated information $\mathbf{t}_v^{l+1}$. After $l$ steps of information aggregation, the final embedding gathers the $l$-hop neighborhood and the structure information. For the graph classification task, readout function $f_{\text{readout}}$ generates a graph level embedding $\mathbf{Z}$ by gathering the embeddings of all nodes in the final layers

$$\mathbf{Z} = f_{\text{readout}}(\{\mathbf{x}_v^{(l)}, v \in \mathcal{V}\}). \tag{3}$$

*2.2.3 SBR Output Layer.* Due to the limited iteration of propagation, GNN cannot effectively capture long-range dependency among items. Therefore, to obtain an effective sequence representation and probability of user preference, existing works propose several strategies to integrate the item representations in the sequence.

## 3 METHODOLOGY

In this section, we present the proposed LLM-based SBR framework LLMGR in detail, which enhances SBR task by integrating language graph structure alignment as illustrated in Fig. 3.

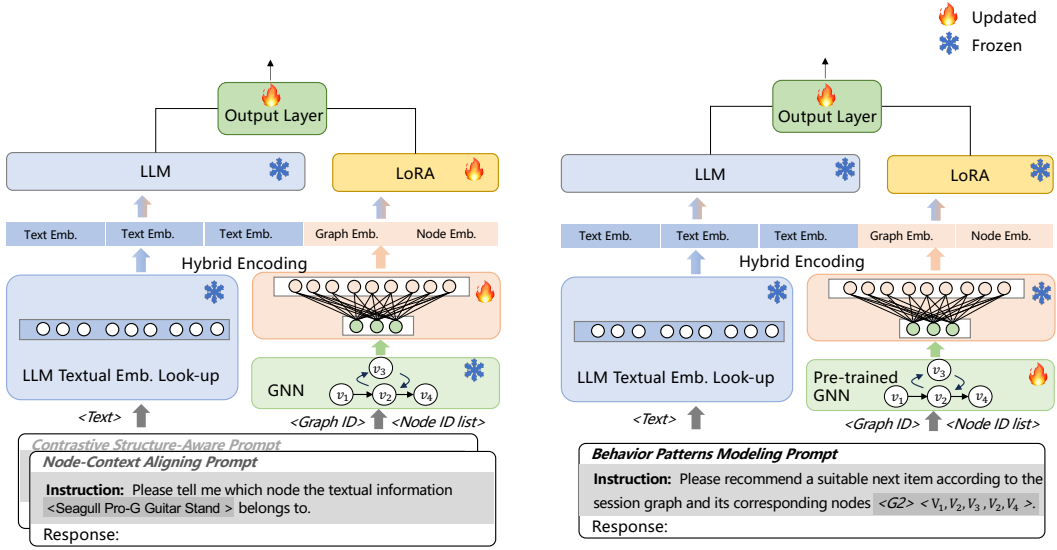

Fig. 3. The Architecture of the LLMGR Framework. The left part describes the auxiliary tuning stage, then the right part illustrates the major instruction tuning stage.

## 3.1 Overview

LLMGR consists of four essential phases: instruction prompts construction, hybrid encoding, LLM output layer, and two-stage instruction tuning. At the stage of multi-task instruction prompts construction, we craft instruction prompts for both auxiliary and major tasks. These prompts are designed to endow the LLM with a nuanced understanding of the session graph's latent user preferences. The major task is designed to understand the inherent structure of the graph, while the auxiliary task is to build interlinkages between nodes and their associated textual information. These instruction prompts incorporate both text and ID components, the latter represents a node from the pre-trained conventional session-based recommender and cannot be directly embedded, therefore we design a hybrid encoding layer to encode them. This layer adeptly converts the prompts into a sequence of embeddings, transforming them into a format that the LLM can process effectively. After that, recommendation results are made by computing the joint probability distribution between the output of LLM and all the candidate items via an item linear transformation. In the tuning phase, due to the significant difference between the auxiliary task and major task, we design a tuning strategy that alternately update the parameters. To sum up, we propose a framework where the structural strengths of graph-based models are effectively combined with the contextual understanding of LLMs, and yield session-based recommender systems capable of delivering highly personalized recommendations.

## 3.2 Instruction Prompt Construction

To simultaneously enable LLM to capture the hidden preferences within session graph, understand the graph structure, and align the nodes with their corresponding textual information, we design a series of prompts to assign language and behavior patterns for tuning LLMs. In the prompts designing stage, the main task is to model behavior pattern model, following by two auxiliary tasks to understand graph structure and align node-text. In these prompts, we incorporate both textual information (template) and placeholders. The placeholders serve the purpose of retrieving embeddings at the hybrid encoding layer 3.3.

*3.2.1 Behavior Patterns Modeling Prompt.* Since our goal is to enhance the performance and generalization ability of the SBR task by leveraging the understanding capability of LLM, behavior pattern modeling has been treated as the major tuning task. We construct personalized recommendation instruction based on user's current session. Then, LLMs are prompted by the instructions with a session graph and corresponding node list to predict the next item that the user is most likely to interact with. The template of instruction tuning is shown in the following:

> **Instruction:** Please recommend a suitable next item according to the session graph and its corresponding nodes <session graph ID> <node ID list>.
>
> **Response:**

In this template, <session graph ID> and <node ID list> are the placeholder that incorporate the graph index and its node, which is used to obtain their embedding in the hybrid encoding layer detail in section 3.3. Taking Fig. 2 as an example, <session graph id> is $G_2$ and the <node list> is $[v_1, v_2, v_3, v_2, v_4]$. By representing the graph in its index and its node, the assembled prompt is shown below:

> **Instruction:** Please recommend a suitable next item according to the session graph and its corresponding nodes list $< G_2 > < V_1, V_2, V_3, V_2, V_4 >$.
>
> **Response:**

This prompt allows for a more comprehensive understanding of the complicated item transition patterns and user behavior preferences, leading to an improved performance in the SBR task.

*3.2.2 Node-Context Aligning Prompt.* To align nodes and the textual context within LLMs more effectively, our main focus is to explore a prompt that can collaborate well with LLMs. By designing a suitable instruction prompt that incorporates both the node of the session graph and the corresponding context, we aim to facilitate seamless integration of node information into LLMs. We present the instruction samples to illustrate the alignment tuning tasks in the following.

> **Instruction:** Please tell me which node the textual information <textual information> belongs to.
>
> **Response:**

The <textual information> refers to the title or description of the target node. Using the node $v_4$ depicted in Fig. 2 as an example, the entire prompt is revealed as follows.

> **Instruction:** Please tell me which node the textual information <Seagull Pro-G Guitar Stand> belongs to.
>
> **Response:**

For this alignment prompt, each node $v \in V$ would be considered. This prompt facilitates a more comprehensive comprehension of nodes, resulting in an increased generalization capability in the SBR task.

*3.2.3 Contrastive Structure-Aware Prompt.* To enhance the understanding capability of large language models on graph structural information, our framework also emphasizes aligning the graph structure with the natural language space. This alignment aims to enable LLMs to effectively comprehend and interpret the structural elements of the graph, maximizing their inherent understanding capabilities. The SBR method primarily focuses on directed graphs as input, therefore we place our emphasis on finding succeeding nodes in the task. The elaborated contrastive structure-aware prompt is shown as follows:

> **Instruction:** Given a target node ID and a candidate node ID list, select the succeeding nodes of the target node <node ID> <candidate node ID list>.
>
> **Response:**

For each session graph $G_i$, the target node is generated from a subset excluding the tail node. The placeholder <candidate node ID list> is randomly arranged using succeeding nodes and randomly sampled negative nodes, which aims to improve the robustness. Using the node $v_3$ depicted in Fig. 2 as the target node, a sample instance is given as follows:

> **Instruction:** Given a target node ID and a candidate node ID list, select the succeeding nodes of the target node $< V_2 > < V_3, V_2, V_6, ..., V_4 >$
>
> **Response:**

This instruction prompts the LLMs to distinguish the positive item from the given candidate, based on the graph structure data information.

## 3.3 Hybrid Encoding Layer

The prompt for our framework consists of text (e.g. template and <textual information>), node ID (e.g. <node ID>, <node ID list>, ), and graph ID (e.g. <session graph ID>). The textual elements of the prompt, such as template, title, and description, are processed by using the LLM's tokenizer and word embedding layer to convert them into tokens and subsequent token embeddings. Despite the fact that LLMs are powerful in modeling natural language and generating reasonable responses, they are unable to comprehend pure ID information due to the lack of textual context. However, in SBR task, the item dependencies within the IDs have been proven to be highly effective in capturing behavioral patterns over a long period. Formally, the ID embeddings $\mathbf{x_v} \in \mathbb{R}^{d_1}$ and graph embedding $\mathbf{z_i} \in \mathbb{R}^{d_1}$ (aforementioned in 2.2.2) are obtained by directly extracting from a pre-trained SBR model, e.g., SRGNN [36], GCSAN [39], HCGR [6], where $\mathbf{\Phi_{SBR}}$ represent their parameters. Since the pre-trained ID embeddings usually have a much lower dimension than the word embeddings in LLM, we employ a linear transformation $f_{in}$ parameterized by $\mathbf{\Phi_{in}}$ to convert the dimension of ID embeddings from $d_1$ into the same dimension as the word embeddings $d_2$, where $d_2$ also much larger than $d_1$. The final inputs are the concatenation of the text embeddings and the ID embeddings in their original positions represented as $\mathbf{E}$. This layer combines the strengths of both text and ID embeddings to enhance the understanding of the input data for LLMs.

## 3.4 LLM Output Layer

After the prompt has been encoded into consecutive embedding $\mathbf{E}$, which can be dealt with LLM Layer. Hence, we integrate the LLaMA-2 [31] and the Low-Rank Adaptation (LoRA) [13] in our LLM layer for instruction tuning. LoRA introduces trainable rank decomposition matrices into pre-trained model layers, notably slashing the necessary trainable parameters for downstream tasks. Then, the output of the LLM layer can be formulated as follows:

$$\mathbf{O} = g_{\mathbf{\Phi_{LLM}+\Phi_{LoRA}}}(\mathbf{E}), \tag{4}$$

where $\mathbf{\Phi_{LLM}}$ represent the parameters of LLM and $\mathbf{\Phi_{LoRA}}$ and Existing LLM-based recommendation approaches, as discussed in previous work [20], typically treat the recommendation task as an open-domain natural language generation task, which align with the nature of LLMs. However, these approaches face challenges in accurately generating recommendations within the defined scope and are limited to generating one recommendation result at a time. Hence, we adopt a Multilayer Perceptron to recommend the probability for each element in the item set $V$ of each given session $S$, which can be formulated as:

$$\hat{\mathbf{y}} = f_{\mathbf{\Phi_{out}}}(\mathbf{O}), \tag{5}$$

where $\mathbf{\Phi_{out}}$ is trainable parameters. The recommendation results are made by computing the joint probability distribution between the output of LLM and all the candidate items via output linear transformation $f_{\mathbf{\Phi_{out}}}$. The output vector $\hat{\mathbf{y}} \in \mathbb{R}^m$, where $m = |V|$. In the tuning stage, the cross entropy loss has been utilized for the learning objective as follows:

$$L = -\sum_{i=1}^{m} y_i log(\hat{y_i}). \tag{6}$$

## 3.5 Two-stage Instruction Tuning Strategy

In this subsection, we will introduce our two-stage instruction tuning and recommendation process, which aim to align LLM with the recommendation task effectively. Due to the significant difference between the auxiliary task and major task, our proposed LLMGR consists of two stages: (1) auxiliary instruction tuning and (2) major instruction tuning. This two-stage optimization enables LLMGR to understand graph structure and align nodes with their context in the former stage, then further capture behavior patterns among session graphs in the latter stage under behavior patterns modeling prompt. Specifically, the auxiliary tuning stage aims to create the association between text and node ID by leveraging node-context alignment prompt 3.2.2 and contrastive structure-aware prompt 3.2.3. During this stage, we tune the parameter of item linear transformation $\mathbf{\Phi_{in}}$ in the hybrid encoding layer, LoRA module $\mathbf{\Phi_{LoRA}}$ and output linear transformation in the LLM output layer, while freezing all other components. After that, the LLM has already acquired the capability to connect node IDs with textual context. However, in the major tuning stage, we focus on capturing behavior preferences within the session graph under behavior patterns modeling prompt 3.2.1, and only unfreeze the parameters $\Phi_{\text{SBR}}$ from the pre-trained SBR method. To be specific, the loss function of LLMGR is the same as Equation 6, where we define the trainable parameters $\Phi$ as:

$$\Phi = \begin{cases} \{\Phi_{\text{LoRA}}, \Phi_{\text{in}}, \Phi_{\text{out}}\}, & \text{auxiliary instruction tuning stage} \\ \{\Phi_{\text{SBR}}, \Phi_{\text{out}}\}, & \text{major instruction tuning stage} \end{cases} \tag{7}$$

## 4 EXPERIMENT

In this section, we provide an overview of the experimental settings, including the datasets, the comparison methods, the evaluation metrics, and the parameter configurations. We then proceed to conduct extensive experiments on three commonly real-world datasets to answer the following research questions:

- RQ1: How does LLMGR perform in SBR scenarios compared with the SOTA methods?
- RQ2: What is the effectiveness and portability of our proposed LLMGR?
- RQ3: How does each component of LLMGR contribute to the performance?
- RQ4: How effective is LLMGR integrating LLM into SBR scenario to alleviate the data sparsity problem?
- RQ5: Can LLMGR present reasonable explanation to predict user preference and get better recommendation results ?

### 4.1 Experimental Settings

*4.1.1 Datasets Description.* We present a summary of the three representative real-world datasets used in our experiments in Table 1. These datasets are well-known Amazon datasets[1], each representing a different category of products, namely Music (Musical Instruments), Beauty (Luxury Beauty), and Pantry (Prime Pantry). To ensure data quality, we follow previous studies [1, 6, 29, 36] and remove inactive users and unpopular items with fewer than five interactions. In addition, we employ the `leave-one-out` strategy, i.e., for each user's interaction sequence, the last item is used as the test data, the second last item is used as the validation data, and the remaining items form the training data.

Table 1. Dataset descriptions.

| Dataset | Music | Beauty | Pantry |
|---|---|---|---|
| #Users | 5,388 | 3,589 | 8,527 |
| #Items | 2,495 | 1,366 | 3,777 |
| #Average Length | 7.44 | 9.99 | 9.35 |
| #Spaisity | 99.70% | 99.92% | 99.75% |
| #Actions | 40,136 | 32,732 | 79,789 |

*4.1.2 Comparison Methods.* To verify the effectiveness of our model, we compare our LLMGR with several baselines as follow:

- **FPMC** [28] - a classical Markov-chain-based method, which considers the latest interaction.

- **CASER** [30] - a CNN-based method that integrates both horizontal and vertical convolutional operations to better capture the high-order interactions within behavior sequences.
- **GRU4Rec** [11] - a representative RNN-based method, which stacks multiple GRU layers and adopts ranking-based loss to the model preference within sessions.
- **NARM** [19] - a hybrid encoder with an attention mechanism to model behavior patterns.
- **STAMP** [25] - an attention model to capture user's temporal interests from historical clicks in a session and relies on self-attention of the last item to represent users' short-term interests.
- **SRGNN** [36] – a GNN-based model for the SBR task, which transforms the session data into a direct unweighted graph and utilizes gated GNN to learn the representation of the item-transitions graph.
- **GCSAN** [39] – a GNN-based model uses gated GNN to extract local context information and then employs the self-attention mechanism to obtain the global representation.
- **NISER** [9] – a GNN-based model using normalized item and session-graph representations to alleviate popularity bias.
- **HCGR** [6] – a GNN-based SBR method enlightened by the powerful representation of non-Euclidean geometry which is proved to be able to reduce the distortion of embedding data onto power-law distribution.

*4.1.3 Evaluation Metrics.* We employ three widely used[7, 8, 36] evaluation metrics, namely $HitRate(H)$, $NDCG(N)$, and $MAP(M)$, to assess the performance of the SBR methods.

To evaluate the ranking ability of the SBR methods, we randomly sample 99 negative items for each positive sample. Additionally, to provide a comprehensive evaluation, we select $K$=5, 10, 20 to calculate the metrics $HitRate@K$, $NDCG@K$, and $MRR@K$. Higher values for all three metrics indicate better performance.

- $HitRate@K$ : If one or more element of the label $y$ is shown in the prediction results $\hat{y}$, we call it a hit. The HitRate is calculated as follows:

$$HitRate@K = \frac{\sum_{s \in S} I\left(y^s \cap \hat{y}^s \neq \Phi\right)}{|S|},\tag{8}$$

where $|\hat{y^s}| = K$ , $I(*)$ denotes the indicator function and $\Phi$ is an empty set. A larger value of HitRate reflects the accuracy of the recommendation results.

- $NDCG@K$ : Normalized Discounted Cumulative Gain ($NDCG$) is a ranking-based metric, that focuses on the order of retrieval results and is calculated in the following way:

$$NDCG@K = \frac{1}{N_k} \sum_{i=1}^{K} \frac{2^{I\left(\hat{y}^s \in y^s\right)} - 1}{\log_2(i+1)},\tag{9}$$

where $N_k$ is a constant to denote the maximum value of $NDCG@K$ given $|\hat{y^s}|$ and $I(*)$ denotes an indicator function. A large NDCG value reflects a higher the ranking position of the expected item.

- $MRR@K$ Mean Reciprocal Rank(MRR) when the r item is not in the higher $K$ position, the reciprocal is set to 0. It is formally given by:

$$MRR@K = \frac{1}{|S|} \sum_{i=1}^{|S|} \frac{1}{\text{rank}_i},\tag{10}$$

where $rank_i$ denotes the position of the item in $\hat{y}^s$. MRR is a normalized ranking that measures the order of recommendation list $y^s$. A large MRR value reflects a higher ranking position of the expected item.

*4.1.4 Implementation Details.* To ensure a fair experimental comparison, we maintain a similar hyperparameter setting for all comparison baselines. Specifically, we set the mini-batch size to 1024, the dropout rate to 0.3, and tune the learning rate from {1e-4, 1e-3, 1e-2, 1e-1}. The embedding size is set to 64, and the maximum sequence length is set to 50 for all models on all datasets. We use the Adam optimizer [18] to update model parameters. For the graph-based methods, we tune the number of graph aggregation layers among {1, 2, 3, 4, 5}.

Table 2. Performance illustration of all comparison methods on all datasets.

| Datasets | Metrics | Comparision Methods | | | | | | | | | Proposed Method | | Improvement |
|---|---|---|---|---|---|---|---|---|---|---|---|---|---|
| | | FPMC | CASER | GRU4Rec | NARM | STAMP | SRGNN | GCSAN | NISER | HCGR | LLMGR-M | LLMGR | |
| Music | HitRate@5 | 0.2548 | 0.2556 | 0.2628 | 0.2782 | 0.2543 | 0.2610 | 0.2905 | 0.2806 | 0.3016 | 0.3047 | **0.3068** | 1.73% |
| | NDCG@5 | 0.1794 | 0.1884 | 0.1867 | 0.2026 | 0.1882 | 0.1901 | 0.2218 | 0.2019 | 0.2226 | 0.2310 | **0.2330** | 4.68% |
| | MRR@5 | 0.1547 | 0.1663 | 0.1618 | 0.1779 | 0.1665 | 0.1669 | 0.1992 | 0.1760 | 0.2001 | 0.2083 | **0.2094** | 4.65% |
| | HitRate@10 | 0.3491 | 0.3536 | 0.3716 | 0.3860 | 0.3478 | 0.3476 | 0.3873 | 0.3870 | 0.3965 | 0.4052 | **0.4085** | 3.02% |
| | NDCG@10 | 0.2097 | 0.2200 | 0.2218 | 0.2374 | 0.2182 | 0.2180 | 0.2530 | 0.2363 | 0.2631 | 0.2633 | **0.2719** | 3.35% |
| | MRR@10 | 0.1671 | 0.1793 | 0.1762 | 0.1921 | 0.1788 | 0.1784 | 0.2120 | 0.1902 | 0.2115 | 0.2216 | **0.2218** | 4.62% |
| | HitRate@20 | 0.4829 | 0.4883 | 0.5117 | 0.5267 | 0.4759 | 0.4731 | 0.5254 | 0.5290 | 0.5224 | 0.5485 | **0.5533** | 4.60% |
| | NDCG@20 | 0.2434 | 0.2541 | 0.2571 | 0.2728 | 0.2505 | 0.2496 | 0.2877 | 0.2723 | 0.2948 | 0.3015 | **0.3051** | 3.51% |
| | MRR@20 | 0.1763 | 0.1887 | 0.1858 | 0.2018 | 0.1876 | 0.1869 | 0.2214 | 0.2002 | 0.2312 | 0.2319 | **0.2355** | 1.89% |
| Beauty | HitRate@5 | 0.4787 | 0.4840 | 0.4723 | 0.4957 | 0.4726 | 0.4712 | 0.5141 | 0.3918 | 0.5156 | 0.5372 | **0.5681** | 10.19% |
| | NDCG@5 | 0.4173 | 0.4237 | 0.4048 | 0.4297 | 0.4247 | 0.4090 | 0.4618 | 0.2841 | 0.4644 | 0.4644 | **0.4924** | 6.03% |
| | MRR@5 | 0.3970 | 0.4038 | 0.3826 | 0.4078 | 0.4088 | 0.3884 | 0.4445 | 0.2482 | 0.4461 | 0.4405 | **0.4673** | 4.77% |
| | HitRate@10 | 0.5531 | 0.5528 | 0.5503 | 0.5662 | 0.5300 | 0.5419 | 0.5756 | 0.4634 | 0.5770 | 0.6342 | **0.6517** | 12.95% |
| | NDCG@10 | 0.4413 | 0.4459 | 0.4298 | 0.4525 | 0.4431 | 0.4319 | 0.4818 | 0.3072 | 0.4830 | 0.4957 | **0.5192** | 7.50% |
| | MRR@10 | 0.4069 | 0.4129 | 0.3927 | 0.4172 | 0.4163 | 0.3978 | 0.4528 | 0.2577 | 0.4528 | 0.4533 | **0.4783** | 5.64% |
| | HitRate@20 | 0.6484 | 0.6509 | 0.6553 | 0.6662 | 0.6119 | 0.6392 | 0.6634 | 0.5807 | 0.6671 | 0.7490 | **0.7501** | 12.34% |
| | NDCG@20 | 0.4653 | 0.4706 | 0.4562 | 0.4775 | 0.4636 | 0.4563 | 0.5038 | 0.3368 | 0.5041 | 0.5247 | **0.5441** | 7.93% |
| | MRR@20 | 0.4135 | 0.4196 | 0.3999 | 0.4239 | 0.4219 | 0.4044 | 0.4587 | 0.2658 | 0.4604 | 0.4613 | **0.4852** | 5.37% |
| Pantry | HitRate@5 | 0.2049 | 0.2151 | 0.2393 | 0.2314 | 0.2083 | 0.2196 | 0.2375 | 0.2335 | 0.2342 | 0.2753 | **0.2845** | 18.87% |
| | NDCG@5 | 0.1340 | 0.1468 | 0.1623 | 0.1547 | 0.1399 | 0.1499 | 0.1656 | 0.1579 | 0.1592 | 0.2006 | **0.2078** | 25.44% |
| | MRR@5 | 0.1108 | 0.1245 | 0.1370 | 0.1297 | 0.1175 | 0.1271 | 0.1421 | 0.1331 | 0.1338 | 0.1764 | **0.1825** | 28.46% |
| | HitRate@10 | 0.3208 | 0.3225 | 0.3547 | 0.3477 | 0.3121 | 0.3221 | 0.3434 | 0.3455 | 0.3489 | 0.3813 | **0.3883** | 9.47% |
| | NDCG@10 | 0.1712 | 0.1813 | 0.1994 | 0.1921 | 0.1732 | 0.1829 | 0.1996 | 0.1940 | 0.1942 | 0.2315 | **0.2413** | 20.88% |
| | MRR@10 | 0.1260 | 0.1385 | 0.1522 | 0.1450 | 0.1311 | 0.1406 | 0.1559 | 0.1479 | 0.1490 | 0.1907 | **0.1963** | 25.89% |
| | HitRate@20 | 0.4728 | 0.4650 | 0.4974 | 0.4930 | 0.4583 | 0.4665 | 0.4784 | 0.4996 | 0.4953 | 0.5215 | **0.5242** | 4.92% |
| | NDCG@20 | 0.2095 | 0.2171 | 0.2352 | 0.2287 | 0.2100 | 0.2192 | 0.2336 | 0.2327 | 0.2335 | 0.2686 | **0.2754** | 17.09% |
| | MRR@20 | 0.1365 | 0.1483 | 0.1619 | 0.1549 | 0.1411 | 0.1504 | 0.1652 | 0.1584 | 0.1585 | 0.1986 | **0.2056** | 24.44% |

Our LLMGR model is implemented based on LLaMA2-7B[2] using the HuggingFace[3] library and accelerated training is performed using DeepSpeed[4]. All experiments are conducted on 2 Nvidia Tesla A100 GPUs. The ID embeddings in our LLMGR model are directly extracted from the pre-trained GCSAN model without modification, as GCSAN has demonstrated its effectiveness and generalization. We employ the AdamW optimizer[24] for model optimization, tune the learning rate among {1e-3, 1e-4, 1e-5}, and set the batch size to 16. Additionally, we utilize a cosine scheduler to adjust the learning rate over steps, and the weight decay is set to 1e-2. In the auxiliary tuning stage, we train for 1 epoch, and in the major tuning stage, we train for 3 epochs on each dataset. In the following section, we investigate the impact of key hyperparameters in greater depth.

## 4.2 Overall Performance (for Q1)

To answer Q1, we display the HitRate, NDCG, and MAP metrics in Table 2. The top-performing values for each metric are emphasized in bold, while the second-best results are underscored. In the final column of this table, we quantify the performance enhancement of LLMGR in comparison to the most competitive methods observed. The results of LLMGR as well as the baselines on three datasets are presented in Table 2. After observation, we can draw the following main findings:

- Deep learning methods (i.e., CASER, GRU4Rec) show their strength in SBR tasks. Their performances are superior to traditional recommendation methods, e.g. FPMC, which prove that adopting a deep learning technology in recommendation systems is a necessary manner since neural networks can model the complex interactions among items.
- The attention-based models, including NARM and STAMP, achieve higher performance metrics compared to non-attention alternatives like GRU4Rec and CASER. The success of attention-based methods stems from their ability to discern shifts in user interests within sessions, effectively pinpointing the primary intent by harnessing attention to synthesizing personal interests from long-term memories or recent short-term behaviors.
- The graph-based methods, i.e., SRGNN, GCSAN, NISER and, HCGR show their superiority compared to graph-free methods, due to the remarkable capacity of graph neural networks to capture complex interaction of user behaviors and describe the coherence of items in a session. NISER presents unstable performance, especially on the Pantry dataset, probably due to its inability to capture long-term dependencies. In contrast, GCSAN and HCGR, by disseminating long-distance information via GNNs and mitigating over-fitting issues, lead the performance among the baseline models.
- Our proposed LLMGR and its variant LLMGR-M significantly outperform all the baseline methods on all three datasets thanks to the powerful ability of LLM. To fairly compare with the conventional SBR methods, we set the variant LLMGR-M, which only trains by the major tuning task without any textual information. The relative improvements of LLMGR in performance over the most competitive baseline methods are about 8.68% on HR@20, 10.71% on NDCG@20, and 11.75% on MRR@20. Taking apart consideration from the extra-textual information, LLMGR-M still outperforms most of the baseline, especially in ranking metrics. The significant improvement fully demonstrates the effectiveness of our framework.

## 4.3 Portability of LLMGR (for Q2)

One of the key innovations of LLMGR is its ability to enhance traditional SBR models through the semantic richness of LLMs. As mentioned earlier, our framework is portable, allowing it to be applied to most existing methods and enhance their performances. As our method requires the use of embeddings, we have only selected deep learning-based methods to adapt to LLMGR. To further illustrate the effectiveness and portability of LLMGR, we apply our framework to the comparison methods. The results between the original methods and the remolded versions on different datasets are plotted in Table 3, with the methods followed by '-L' denoting their variant under our proposed LLMGR. However, our framework is specifically designed for graph-based SBR methods. To adapt our framework to graph-free methods, we conceal the '<session graph ID>' placeholder from the behavior patterns modeling prompt during the major tuning task. From Table 3, we can draw the following conclusions:

- LLMGR works well with various SBR models, not only with the GCSAN. As observed, the variant methods outperform their original versions with average improvements of 8.58%, and 17.09% on Music, and Beauty, respectively, which demonstrates the effectiveness of LLMGR.
- The improvements on the originally less effective models are even more significant, such as STAMP and GRU4Rec. Besides, we find that simpler models(e.g. GRU4Rec, STAMP) can outperform most baseline SBR models under LLMGR, which indicates that LLMGR effectively provides additional information to the SBR models.
- Under our LLMGR framework, the baseline demonstrates significant improvements across all metrics, especially at smaller K values. Furthermore, ranking performance metrics such as NDCG and MRR show considerably larger improvements compared to HitRate. This suggests that our proposed LLMGR effectively captures user preferences, resulting in more accurate and reliable recommendation outcomes.

Table 3. Portability of LLMGR on Music and Beauty datasets.

| Dataset | Method | HitRate@5 | NDCG@5 | MRR@5 | HitRate@10 | NDCG@10 | MRR@10 | HitRate@20 | NDCG@20 | MRR@20 |
|---------|--------|-----------|--------|-------|------------|---------|--------|------------|---------|--------|
| Music | CASER | 0.2556 | 0.1884 | 0.1663 | 0.3536 | 0.2200 | 0.1793 | 0.4883 | 0.2541 | 0.1887 |
| | CASER-L | 0.2678 | 0.1986 | 0.1759 | 0.3558 | 0.2268 | 0.1874 | 0.4839 | 0.2590 | 0.1961 |
| | **Improvement** | **4.79%** | **5.41%** | **5.74%** | **0.63%** | **3.09%** | **4.49%** | **-0.91%** | **1.92%** | **3.93%** |
| | GRU4Rec | 0.2628 | 0.1867 | 0.1618 | 0.3716 | 0.2218 | 0.1762 | 0.5117 | 0.2571 | 0.1858 |
| | GRU4Rec-L | 0.2919 | 0.2211 | 0.1979 | 0.3920 | 0.2533 | 0.2111 | 0.5236 | 0.2864 | 0.2201 |
| | **Improvement** | **11.09%** | **18.43%** | **22.31%** | **5.49%** | **14.21%** | **19.78%** | **2.32%** | **11.41%** | **18.43%** |
| | NARM | 0.2782 | 0.2026 | 0.1779 | 0.3860 | 0.2374 | 0.1921 | 0.5267 | 0.2728 | 0.2018 |
| | NARM-L | 0.2892 | 0.2146 | 0.1902 | 0.3846 | 0.2456 | 0.2030 | 0.5135 | 0.2781 | 0.2119 |
| | **Improvement** | **3.94%** | **5.93%** | **6.93%** | **-0.38%** | **3.45%** | **5.65%** | **-2.50%** | **1.94%** | **5.00%** |
| | STAMP | 0.2543 | 0.1882 | 0.1665 | 0.3478 | 0.2182 | 0.1788 | 0.4759 | 0.2505 | 0.1876 |
| | STAMP-L | 0.3009 | 0.2300 | 0.2067 | 0.3862 | 0.2574 | 0.2179 | 0.5161 | 0.2902 | 0.2269 |
| | **Improvement** | **18.32%** | **22.22%** | **24.13%** | **11.05%** | **17.98%** | **21.89%** | **8.46%** | **15.85%** | **20.94%** |
| | SRGNN | 0.2610 | 0.1901 | 0.1669 | 0.3476 | 0.2180 | 0.1784 | 0.4731 | 0.2496 | 0.1869 |
| | SRGNN-L | 0.2699 | 0.2018 | 0.1795 | 0.3729 | 0.2349 | 0.1931 | 0.5009 | 0.2672 | 0.2019 |
| | **Improvement** | **3.41%** | **6.17%** | **7.56%** | **7.26%** | **7.76%** | **8.25%** | **5.88%** | **7.08%** | **8.00%** |
| | GCSAN | 0.2905 | 0.2218 | 0.1992 | 0.3873 | 0.2530 | 0.2120 | 0.5254 | 0.2877 | 0.2214 |
| | GCSAN-L | 0.3068 | 0.2330 | 0.2094 | 0.4085 | 0.2719 | 0.2218 | 0.5533 | 0.3051 | 0.2355 |
| | **Improvement** | **5.62%** | **5.08%** | **5.12%** | **5.46%** | **7.47%** | **4.62%** | **5.30%** | **6.08%** | **6.39%** |
| | NISER | 0.2806 | 0.2019 | 0.1760 | 0.3870 | 0.2363 | 0.1902 | 0.5290 | 0.2723 | 0.2002 |
| | NISER-L | 0.3048 | 0.2322 | 0.2084 | 0.4063 | 0.2649 | 0.2218 | 0.5380 | 0.2981 | 0.2309 |
| | **Improvement** | **8.60%** | **14.98%** | **18.38%** | **4.99%** | **12.08%** | **16.61%** | **1.72%** | **9.50%** | **15.30%** |
| | HCGR | 0.3016 | 0.2226 | 0.2001 | 0.3965 | 0.2631 | 0.2115 | 0.5224 | 0.2948 | 0.2312 |
| | HCGR-L | 0.3171 | 0.2358 | 0.2128 | 0.4256 | 0.2836 | 0.2272 | 0.5534 | 0.3151 | 0.2479 |
| | **Improvement** | **5.15%** | **5.90%** | **6.36%** | **7.33%** | **7.80%** | **7.41%** | **5.94%** | **6.90%** | **7.23%** |
| Beauty | CASER | 0.4840 | 0.4237 | 0.4038 | 0.5528 | 0.4459 | 0.4129 | 0.6509 | 0.4706 | 0.4196 |
| | CASER-L | 0.5467 | 0.4774 | 0.4546 | 0.6333 | 0.5053 | 0.4661 | 0.7297 | 0.5297 | 0.4727 |
| | **Improvement** | **12.95%** | **12.66%** | **12.57%** | **14.57%** | **13.32%** | **12.86%** | **12.11%** | **12.56%** | **12.65%** |
| | GRU4Rec | 0.4723 | 0.4048 | 0.3826 | 0.5503 | 0.4298 | 0.3927 | 0.6553 | 0.4562 | 0.3999 |
| | GRU4Rec-L | 0.5497 | 0.4800 | 0.4570 | 0.6291 | 0.5055 | 0.4675 | 0.7295 | 0.5309 | 0.4745 |
| | **Improvement** | **16.40%** | **18.56%** | **19.45%** | **14.33%** | **17.62%** | **19.04%** | **11.31%** | **16.37%** | **18.64%** |
| | NARM | 0.4957 | 0.4297 | 0.4078 | 0.5662 | 0.4525 | 0.4172 | 0.6662 | 0.4775 | 0.4239 |
| | NARM-L | 0.5542 | 0.4828 | 0.4592 | 0.6411 | 0.5105 | 0.4704 | 0.7434 | 0.5363 | 0.4775 |
| | **Improvement** | **11.80%** | **12.34%** | **12.61%** | **13.24%** | **12.81%** | **12.76%** | **11.59%** | **12.31%** | **12.64%** |
| | STAMP | 0.4726 | 0.4247 | 0.4088 | 0.5300 | 0.4431 | 0.4163 | 0.6119 | 0.4636 | 0.4219 |
| | STAMP-L | 0.5375 | 0.4716 | 0.4498 | 0.6325 | 0.5024 | 0.4626 | 0.7345 | 0.5280 | 0.4695 |
| | **Improvement** | **13.74%** | **11.04%** | **10.03%** | **19.35%** | **13.38%** | **11.12%** | **20.04%** | **13.88%** | **11.30%** |
| | SRGNN | 0.4712 | 0.4090 | 0.3884 | 0.5419 | 0.4319 | 0.3978 | 0.6392 | 0.4563 | 0.4044 |
| | SRGNN-L | 0.5497 | 0.4800 | 0.4570 | 0.6291 | 0.5055 | 0.4675 | 0.7295 | 0.5309 | 0.4745 |
| | **Improvement** | **16.68%** | **17.34%** | **17.65%** | **16.09%** | **17.06%** | **17.52%** | **14.12%** | **16.35%** | **17.31%** |
| | GCSAN | 0.5141 | 0.4618 | 0.4445 | 0.5756 | 0.4818 | 0.4528 | 0.6634 | 0.5038 | 0.4587 |
| | GCSAN-L | 0.5681 | 0.4924 | 0.4673 | 0.6517 | 0.5192 | 0.4783 | 0.7501 | 0.5441 | 0.4852 |
| | **Improvement** | **10.51%** | **6.62%** | **5.14%** | **13.21%** | **7.78%** | **5.65%** | **13.06%** | **8.01%** | **5.78%** |
| | NISER | 0.3918 | 0.2841 | 0.2482 | 0.4634 | 0.3072 | 0.2577 | 0.5807 | 0.3368 | 0.2658 |
| | NISER-L | 0.5035 | 0.4339 | 0.4110 | 0.5896 | 0.4617 | 0.4225 | 0.6971 | 0.4889 | 0.4299 |
| | **Improvement** | **28.52%** | **52.73%** | **65.59%** | **27.24%** | **50.28%** | **63.92%** | **20.06%** | **45.17%** | **61.76%** |
| | HCGR | 0.5156 | 0.4644 | 0.4461 | 0.5770 | 0.4830 | 0.4528 | 0.6677 | 0.5041 | 0.4604 |
| | HCGR-L | 0.5751 | 0.5016 | 0.4761 | 0.6568 | 0.5237 | 0.4864 | 0.7513 | 0.5593 | 0.4870 |
| | **Improvement** | **11.54%** | **8.03%** | **6.73%** | **13.82%** | **8.42%** | **7.43%** | **12.53%** | **10.94%** | **5.77%** |

- The performances of all kinds of models, not only graph-based models, improve significantly on the three datasets, which demonstrates that the world knowledge and context comprehension abilities of LLM could enhance item understanding and user modeling. Such abilities in session-based recommendation data are essential for predicting users' nuanced behavior, while such information is just ignored by the traditional SBR models.

Table 4. The performance with different instruction prompts on Music and Beauty datasets.

| Dataset | Method | HitRate@5 | NDCG@5 | MRR@5 | HitRate@10 | NDCG@10 | MRR@10 | HitRate@20 | NDCG@20 | MRR@20 |
|---------|--------|-----------|--------|-------|------------|---------|--------|------------|---------|--------|
| Music | LLMGR-M(SRGNN) | 0.2676 | 0.2018 | 0.1801 | 0.3653 | 0.2334 | 0.1931 | 0.4952 | 0.2661 | 0.2020 |
| | LLMGR(SRGNN) | 0.2699 | 0.2018 | 0.1795 | 0.3729 | 0.2349 | 0.1931 | 0.5009 | 0.2672 | 0.2019 |
| | **Improvement** | **0.83%** | **-0.01%** | **-0.33%** | **2.08%** | **0.67%** | **-0.01%** | **1.16%** | **0.41%** | **-0.08%** |
| | LLMGR-M(GCSAN) | 0.3018 | 0.2300 | 0.2063 | 0.3996 | 0.2615 | 0.2193 | 0.5254 | 0.2932 | 0.2279 |
| | LLMGR(GCSAN) | 0.3047 | 0.2310 | 0.2083 | 0.4052 | 0.2633 | 0.2216 | 0.5485 | 0.3015 | 0.2319 |
| | **Improvement** | **0.97%** | **0.45%** | **0.93%** | **1.39%** | **0.69%** | **1.05%** | **4.40%** | **2.82%** | **1.74%** |
| | LLMGR-M(NISER) | 0.3036 | 0.2276 | 0.2025 | 0.4066 | 0.2607 | 0.2161 | 0.5340 | 0.2927 | 0.2248 |
| | LLMGR(NISER) | 0.3048 | 0.2322 | 0.2084 | 0.4063 | 0.2649 | 0.2218 | 0.5380 | 0.2981 | 0.2309 |
| | **Improvement** | **0.37%** | **2.03%** | **2.91%** | **-0.09%** | **1.60%** | **2.66%** | **0.76%** | **1.83%** | **2.71%** |
| Beauty | LLMGR-M(SRGNN) | 0.5102 | 0.4437 | 0.4217 | 0.5993 | 0.4725 | 0.4336 | 0.7155 | 0.5018 | 0.4416 |
| | LLMGR(SRGNN) | 0.5497 | 0.4800 | 0.4570 | 0.6291 | 0.5055 | 0.4675 | 0.7295 | 0.5309 | 0.4745 |
| | **Improvement** | **7.76%** | **8.19%** | **8.37%** | **4.97%** | **6.99%** | **7.82%** | **1.95%** | **5.81%** | **7.45%** |
| | LLMGR-M(GCSAN) | 0.5372 | 0.4644 | 0.4405 | 0.6342 | 0.4957 | 0.4533 | 0.7490 | 0.5247 | 0.4613 |
| | LLMGR(GCSAN) | 0.5681 | 0.4924 | 0.4673 | 0.6517 | 0.5192 | 0.4783 | 0.7501 | 0.5441 | 0.4852 |
| | **Improvement** | **5.76%** | **6.02%** | **6.10%** | **2.77%** | **4.76%** | **5.52%** | **0.15%** | **3.70%** | **5.17%** |
| | LLMGR-M(NISER) | 0.4865 | 0.4153 | 0.3918 | 0.5731 | 0.4431 | 0.4032 | 0.6879 | 0.4722 | 0.4112 |
| | LLMGR(NISER) | 0.5035 | 0.4339 | 0.4110 | 0.5896 | 0.4617 | 0.4225 | 0.6971 | 0.4889 | 0.4299 |
| | **Improvement** | **3.49%** | **4.48%** | **4.90%** | **2.87%** | **4.20%** | **4.79%** | **1.34%** | **3.53%** | **4.56%** |

## 4.4 Ablation Studies (for Q3)

*4.4.1 Effect of Multi-task Instruction Prompt.* Our proposed LLMGR consists of two tuning stages. The auxiliary instruction tuning stage aims to endow the LLM to understand the structure of the session graph and establish text-node associations. The major tuning task focuses on capturing the behavioral patterns within session graphs under LLM. To answer Q3, we conduct ablation experiments with a simplified version of LLMGR, e.g., LLMGR-M. This variant removes the node-context aligning prompt and contrastive structure-aware prompt (auxiliary tuning stage), only tuning under major tuning task. Furthermore, to gain a deeper insight into our tailored prompt, we adapt the selected comparison pre-trained graph-based SBR (e.g., SRGNN, GCSAN, NISER) to conduct this experiment on the Music and Beauty dataset. The experimental results are shown in Table 4, we can draw the following observations:

- After taking apart the auxiliary tuning prompt from the LLMGR framework, which means removing the two alignment tasks for LLM under the tuning stage. This operation may disable the LLMGR from understanding the graph structure and realize text of nodes. As expected, this modification leads to a marked decline in performance across nearly all metrics, with a more pronounced drop observed at lower K values. Moreover, the degradation is more substantial in ranking-based metrics like NDCG and MRR than in Hit Rate. These observations indicate that LLMGR is particularly adept at discerning user preferences, which translates to enhanced accuracy and dependability in the recommendations it generates. Besides, the world knowledge and context comprehension abilities of LLMs could enhance item understanding and user modeling.
- By comparing LLMGR to LLMGR-M among all pre-trained SBR methods, it is obvious that after removing the auxiliary tuning stage, the average performance drops by 2.04%, 1.65% and 1.42% in terms of HitRate@20, NDCG@20, and MRR@20 on Music dataset, respectively, and the corresponding average decrease rates are 1.13%, 4.16% and 5.40% on Beauty.

*4.4.2 Effect of Two-stage Tuning Strategy.* In the above section, we have demonstrated the effectiveness of our multi-task instruction prompt. In this section, we will examine the effects of our two-stage tuning strategy on the LLMGR's efficacy. This two-stage optimization process first equips the LLMGR with the ability to decode graph structures and align nodes with their respective contexts, then further model behavioral patterns within session graphs using behavior pattern modeling prompts. In this subsection, we further explore the following tuning strategies:

- **Single-stage Tuning (LLMGR-O):** Here, we employ only one tuning approach, directly tuning all the trainable parameters, including $\Phi_{LoRA}$, $\Phi_{in}$, $\Phi_{out}$ and $\Phi_{SBR}$. This means that the entire range of prompts implies that all kinds of prompts(e.g., behavior modeling and graph comprehension) will be trained together.
- **Two-stage Tuning Without Freeze (LLMGR-F):** We first fine-tune LLMGR under the auxiliary task and then proceed to fine-tune the major task. It is important to note that, in this variant, none of the parameters will be frozen.

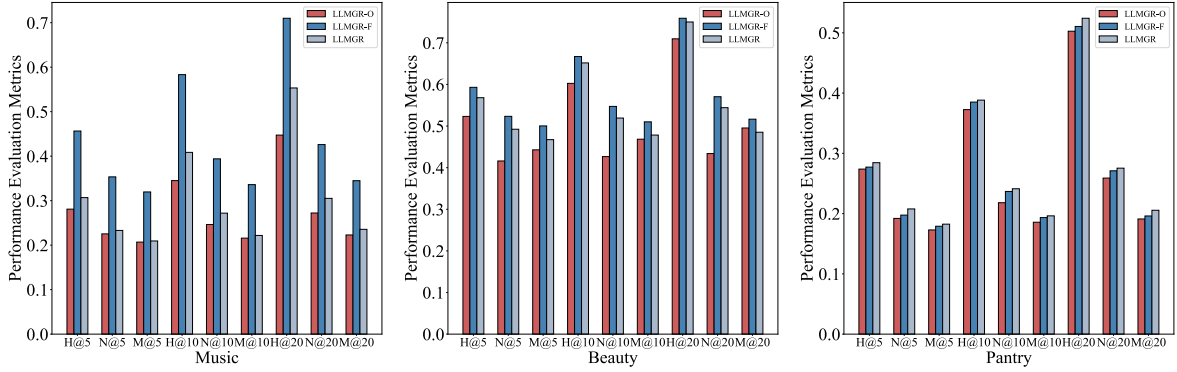

Fig. 4. The performance with different tuning strategies on three datasets.

We compare these methods in terms of their overall performance, as shown in Fig. 4, and we can draw the following observations:

- LLMGR outshines both LLMGR-O and LLMGR-F across all scenarios within the two datasets. The success of LLMGR can be linked to its strategy of segregating distinct prompt types during separate tuning stages, which more effectively discerns the shared characteristics and unique aspects of various tasks. The observed improvements may be ascribed to the two-stage tuning strategy's ability to mitigate interference between semantic information and behavioral pattern modeling, thereby enhancing the overall recommendation performance. Furthermore, by allocating distinct prompts to different stages of the tuning process, LLMGR ensures a focused optimization for each task. During the initial stage, the model concentrates on grasping the semantics of the items and the user's immediate context. This forms a solid foundation for the subsequent stage, where LLMGR hones in on intricate behavioral patterns, utilizing the groundwork laid in the first stage to yield a more nuanced understanding of user interactions.
- An in-depth examination of the outcomes demonstrates that our proposed LLMGR method uniformly surpasses the other tuning strategy variant across both datasets. This consistent edge in performance highlights the efficacy of our two-stage tuning strategy.

## 4.5 Effectiveness Analysis (for Q4)

*4.5.1 Analysis on Session Lengths.* This section aims to analyze the performance of various recommendation models in handling sessions with different lengths, specifically in the context of the Music and Beauty datasets. In order to conduct a comparative analysis, the sessions are classified into two distinct categories based on their length. "Short" sessions, which consist of seven items or fewer. The choice of seven as a threshold is strategic—it aligns closely with the average session length that has been observed across all datasets examined in the study.

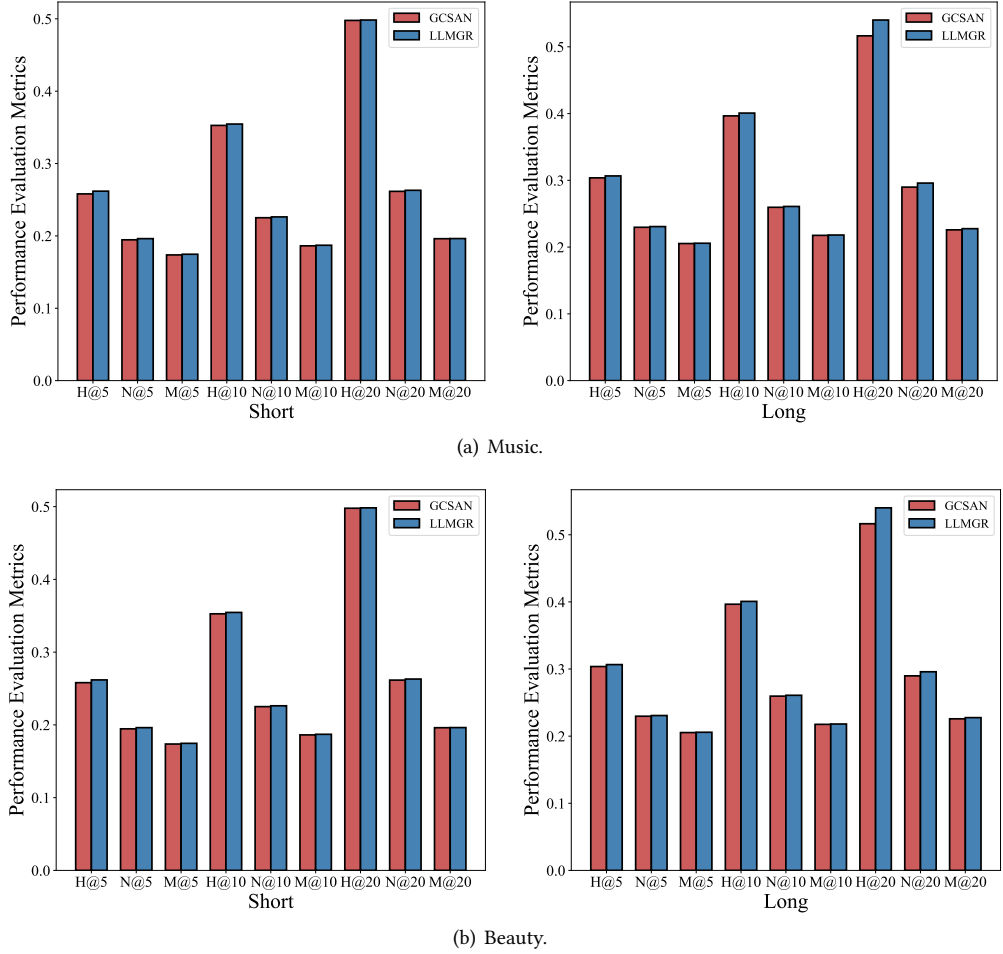

(a) Music.

(b) Beauty.

Fig. 5. Performance comparison on different session lengths evaluated in all metrics on two datasets.

By defining "short" in this way, the researchers are focusing on sessions that are at or below the average length, looking to understand how well different models can handle sessions that may not provide a substantial amount of user-item interaction data. "Long" sessions, are defined as containing more than seven items. These sessions exceed the average length and thus present a different kind of challenge for recommendation systems—they typically provide more interaction data within a single session, which could be beneficial for making accurate recommendations but might also introduce more complexity due to the variety of items and user preferences reflected in the longer session data. By segmenting the sessions in this manner, the study intends to shed light on the adaptability and efficiency of different recommender systems models when confronted with varying amounts of information. The performance of the models can be influenced by the amount of data they have to work with—too little data might not provide enough insights for accurate recommendations, while too much data might make it difficult to identify the most relevant items for the user. The performance metrics for each model across both session lengths are detailed in Fig. 5. Based on the results shown in Fig. 5, we have observations as follows:

- Our proposed LLMGR and its variants LLMGR-M perform stably on two datasets with different session lengths. It demonstrates the superior ability of the proposed method and the adaptability of LLM in SBR scenario.
- The superior performance of LLMGR over LLMGR-M, indicates a robustness in the core design of LLMGR. This robustness allows it to consistently outperform not just its variant, but potentially other SBR methods as well, across different session lengths. The fact that LLMGR excels in both long and short sessions is noteworthy, as it implies that the model is not merely effective in one scenario but is genuinely versatile, dealing well with both rich and sparse data environments. The auxiliary tuning task mentioned suggests that LLMGR is equipped with a mechanism that helps contextualize each node within the graph structure of the session data. By aligning the node with its contextual information, LLMGR can leverage additional signals for representation learning. This provides a more nuanced and semantically rich representation of user preferences, which is particularly beneficial in capturing the underlying patterns in user behavior that may not be immediately apparent from interaction data alone.

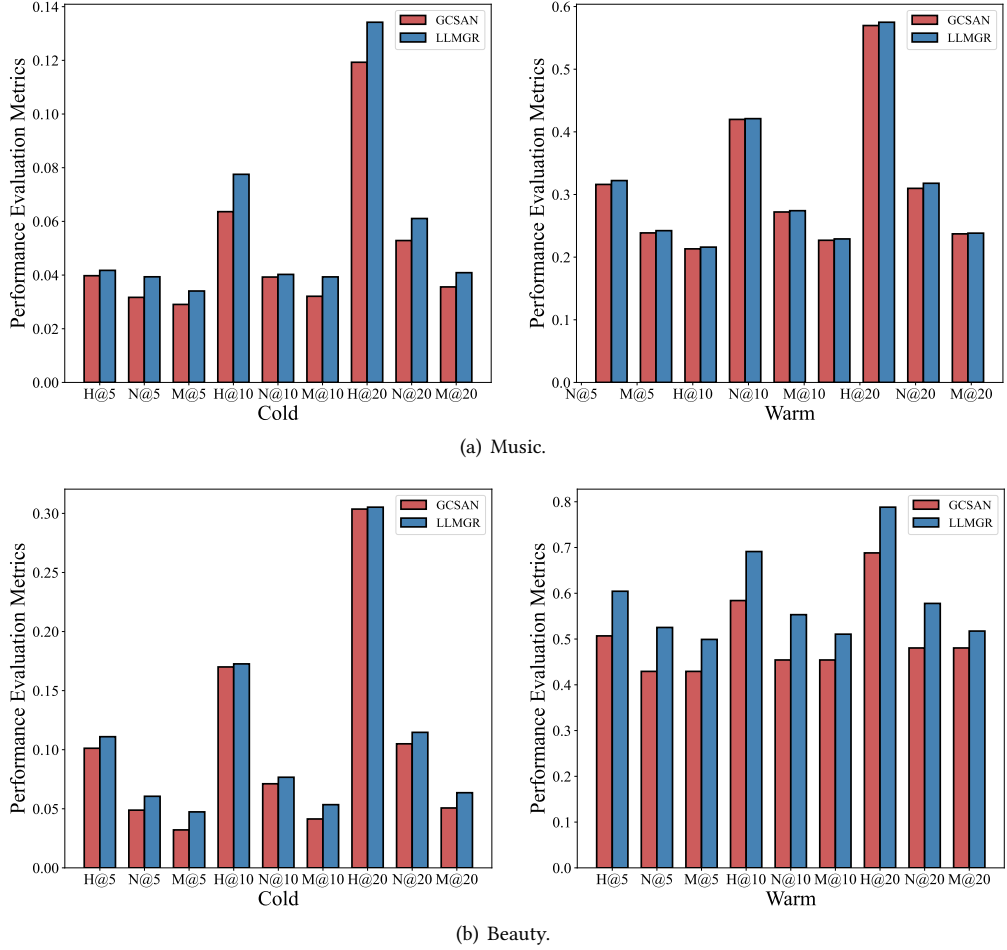

(a) Music.

(b) Beauty.

Fig. 6. Performance comparison on cold and warm items evaluated in all metrics on two datasets.

*4.5.2 Analysis on Cold-start Scenarios.* In the field of recommender systems, the cold-start problem is indeed a significant hurdle. It primarily affects new users or items that have recently been added to the system's database. For new users, the system struggles to make personalized recommendations because it lacks data on their preferences. For new items, the system may not have enough interaction data to establish where or to whom these items might appeal. To test the effectiveness of LLMGR, an experiment was set up where the testing set was divided into warm and cold subsets. The 'warm' subset acts as a benchmark for the performance of LLMGR in ideal conditions, where user-item interactions are abundant and the system has had ample opportunity to learn from these interactions. A strong performance in the warm subset would indicate that LLMGR is at least on par with traditional recommendation techniques in a data-rich environment. The 'cold' subset is the true test of LLMGR's innovation. Success in this subset would demonstrate that LLMGR can effectively alleviate the cold-start problem and suggest that it has mechanisms in place to derive meaningful recommendations from limited interaction data. The observations made from Figure 6

- The significant results obtained from the cold scenario, where each item has limited interactions and it is challenging to capture behavioral preferences, further underscore the robustness of our proposed LLMGR when handling extremely sparse data. Additionally, it also showcases the advantages of the LLM-based method in cold scenarios, highlighting how traditional SBR methods lack the proficiency to effectively handle such scenarios.
- When comparing the performance in warm and cold scenarios, the performance gains are notably more pronounced in the cold scenario. This could be attributed to the capability of LLMs, which enhances the performance of LLMGR, enabling it to achieve satisfactory results with limited data feat that traditional recommendation methods often struggle to accomplish.

Overall, the evaluation from the separation of testing data into 'warm' and 'cold' subsets enables a nuanced evaluation of LLMGR's performance and highlights its potential strengths in tackling one of the most persistent issues in recommender systems.

## 5 CASE STUDY (FOR Q5)

### 5.1 Emebdding Visualization

As demonstrated in section 4.3, our framework is portable and allows it to be applied to most existing methods and enhance their performances. To further investigate the effects of pre-trained SBR methods and it after tuning under our proposed LLMGR. Following previous work [15, 33], we employ Principal Component Analysis (PCA) to visualize the node embeddings. "GCSAN" and "SRGNN" represent item embedding from the corresponding pre-trained SBR method, while "LLMGR" represents the item embedding under our proposed framework. As shown in Fig. 7,8, the nodes of all pre-trained models are concentrated in the middle, with poor differentiation. After tuning under LLMGR, the representation distribution of the nodes becomes more uniform, which is beneficial for mitigating interference between semantic information and behavioral pattern modeling.

## 6 RELATED WORKS

### 6.1 Session-based Recommendation

Early approaches to session-based recommendation are heavily reliant on Markov chains (MC). For instance, FPMC [28] merges matrix factorization with MC to learn both the general preferences and the short-term interests of users for basket recommendation. Meanwhile, Fossil [10] integrates similarity-based methods with MC to offer personalized session-based recommendations. However, a key drawback of MC-based models is that they have been struggling to capture long-range dependencies, under the assumption that future actions are only dependent on the most recent past state. In recent years, Recurrent Neural Networks have been adopted to model temporal dependencies. Pioneering the RNN-based session recommendation, GRU4Rec [11] utilizes the Gated Recurrent Unit (GRU) to capture long-term dependencies across sessions, significantly outperforming MC-based methods with a novel pairwise ranking loss. Building on the success of GRU4Rec, MV-RNN [4] enhances recommendations by incorporating multimodal data, and ROM [44] employs an interactive self-attention mechanism for rating prediction. Nevertheless, RNNs inherently assumes a fixed sequential dependence between adjacent items, resulting in potential inaccuracies and noise in session data, especially in scenarios like music recommendation. Convolutional Neural Networks have also been leveraged to discern patterns in session-based recommendation. CNN-based models [30] apply convolutional filters to capture varying orders of user behavior. More recently, attention mechanism-based models [32] have shown exceptional performance. For example, a hybrid encoder with an attention mechanism was used by Li et al. to model both sequential behavior and session interests [19]. STAMP [23] introduces a short-term attention priority model that effectively captured both the long-term session context and the user's immediate interests. Advanced session-based recommendation models employ attention mechanisms to identify user behavior patterns over long sequences. However, discerning both the implicit and explicit relations between adjacent behaviors remains challenging.

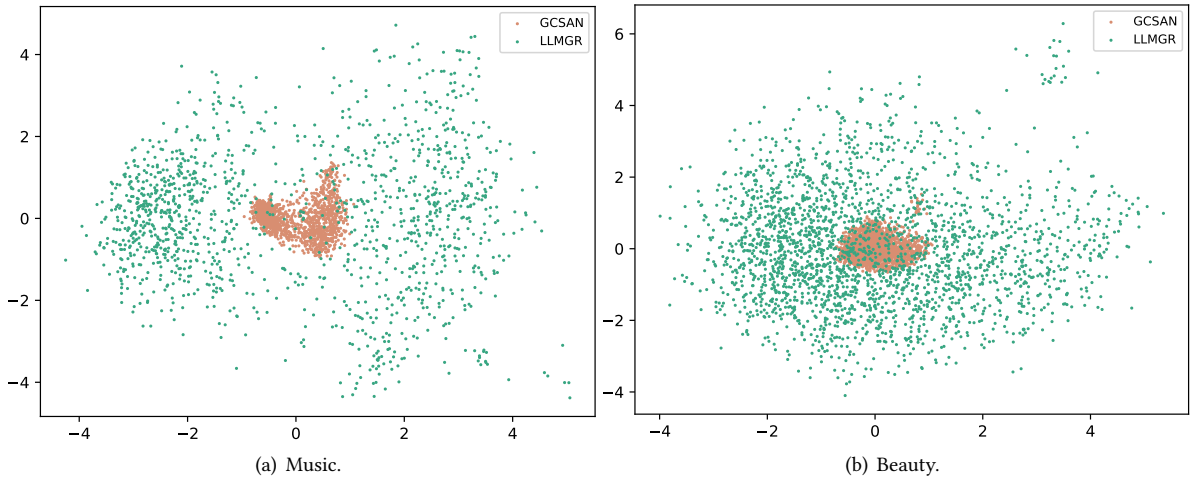

Fig. 7. The embedding visualization of LLMGR and comparison method (GCSAN) on music and beauty dataset.

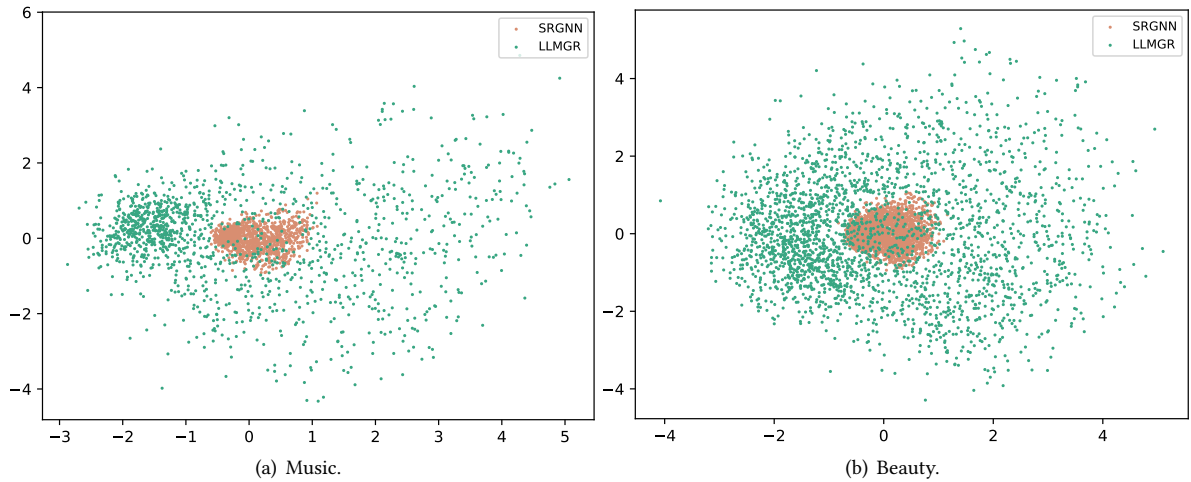

Fig. 8. The embedding visualization of LLMGR and comparison method (SRGNN) on music and beauty dataset.

Graph Neural Networks (GNNs) are adept at uncovering such relationships [14, 35] and can capture complex user behavior interactions. SRGNN [36] and GCSAN [39] represent two such methods that build directed graphs for sequences and employ self-attention to enhance representation learning, respectively. Further, Wu et al. focuses on capturing user history within sessions using a dot-attention mechanism [37], while FGNN [27] proposes a weighted attention layer for learning embeddings. Memory models have also been utilized to encompass both long-term and short-term behaviors [26]. While GNN-based methods have seen significant success, they predominantly have significant challenges in understanding the content of items.

## 6.2 Large Language Models for Recommendation

As Large Language Models (LLMs) have driven impressive innovations in artificial intelligence recently, researchers have been trying to leverage their strong capabilities in language semantic understanding on recommender systems. At first, researchers focus on using prompting or in-context learning. ChatREC[5] propose the idea of an interactive recommend system by utilizing ChatGPT prompt to inject user information and instruct LLMs to generate recommend results. Similar to the above concept, Wang et al. [34] incorporate research into a 3-step prompting that guides GPT-3 to carry sub-tasks that capture the user's preferences and generate a ranked list recommendation. Liu et al.[22] propose a general recommendation prompt construction framework, relying only on the prompts themselves to convert recommendation tasks into natural language tasks.

However, since the underlying semantics of natural language and recommender systems are incompatible, and LLMs have not been pre-trained for recommender systems, researchers have made efforts to address this problem. By fine-tuning LLMs with empirical recommendation data, Zhang et al. [40] and Bao et al.[3] have achieved superior model performances. Based on comprehensive research, Kang et al.[17] discovers the importance of user interaction data and proposes to format the user historical interactions as prompts. Moreover, some concurrent research works have also shed light on bridging the gap between natural language and recommender systems. BIGRec[2] integrates collaborative information and utilizes a two-step grounding framework to generate recommendations with LLM. TransRec [21] employs multi-facet identifiers, combining ID, title, and attributes to balance item distinctiveness and semantics. CoLLM [42] captures collaborative information through an external traditional model and maps it to the input token embedding space of LLM, forming collaborative embeddings for LLM usage[20]. The above works have made great progress in adapting LLMs to recommendation systems, but have problems in providing personalized recommendations due to lack of sophisticated task training.

## 7 CONCLUSIONS

SBR systems play a crucial role in capturing users' dynamic preferences by analyzing the session of their interactions. The SOTA in this domain has been significantly advanced by the application of GNNs. These graph-based SBR methods excel at capturing the intricate patterns of user behaviors and preferences, which is especially crucial in scenarios with limited interaction data. Despite these notable strides, existing SBR methods focus on user interactions while often sidelining the rich textual content that accompanies users and items. This has limited their ability to grasp the full context and nuances behind user preferences. With the advent of LLMs, there is now an opportunity to transcend these limitations by integrating their sophisticated natural language understanding capabilities with SBR systems.

In this work, we have presented LLMGR, a novel framework that synergizes LLMs with graph-based SBR approaches. LLMGR is designed to overcome the challenges posed by the integration of these two powerful technologies. It addresses two central challenges: first, the expression of graph-based SBR tasks in natural language, and second, the alignment of textual information with graph nodes to effectively capture user preferences. Our approach successfully bridges the gap between the graph-structured data of SBR tasks and the sequential processing strengths of LLMs. Through an innovative two-stage tuning process, LLMGR provides a means to associate text with graph nodes and subsequently capture the dynamic behavioral patterns of users. The approach not only enhances the performance of session-based recommendations but also pioneers a new direction for future research at the convergence of complex graph modeling and advanced language processing models. The extensive experiments conducted on three real-world datasets affirm the efficacy of LLMGR, showcasing its superiority over several competitive baselines. This validates the potential of our approach to push the boundaries of SBR systems by enriching them with the depth and breadth of understanding that LLMs offer. In future work, we will explore how to extend the current approach in multi-domain recommendations, so that it can support more flexible interaction with users.